\documentclass[10pt,a4paper, twocolumn]{article}
\usepackage[utf8]{inputenc}
\usepackage{amsmath}
\usepackage{amsfonts}
\usepackage{amssymb}
\usepackage{graphicx}
\usepackage{indentfirst}
\usepackage{cite}
\usepackage{textcomp}
\usepackage{standalone}
\usepackage{cuted}

\DeclareMathOperator\arctanh{arctanh}
\title{Electrical conductivity of quasi-2D foams}
\author{Pavel Yazhgur$^{\ast}$ \and Cl\'ement Honorez \and Wiebke Drenckhan  \and Dominique Langevin \and Anniina Salonen$^{\ast}$ }
\date{}

\begin{document}
\renewcommand{\thefootnote}{\fnsymbol{footnote}}
\maketitle

\section{Abstract}
\label{sec:Abstract}

Quasi-2D foams consist of monolayers of bubbles squeezed between two narrowly spaced plates. These simplified foams have served successfully in the past to shed light on numerous issues in foam physics. Here we consider the electrical conductivity of such model foams. We compare experiments to a model which we propose, and which successfully relates the structural and the conductive properties of the foam over the full range of the investigated liquid content. We show in particular that in the case of quasi-2D foams the liquid in the nodes needs to be taken into account even at low liquid content. We think that these results may provide new approaches for the characterisation of foam properties and for the in-situ characterisation of the liquid content of foams in confining geometries, such as microfluidics.

\section{Introduction}
\label{sec:Introduction}

\footnotetext{$^{\ast}$ Authors to whom correspondence should be adressed: pavel.yazhgur@u-psud.fr and anniina.salonen@u-psud.fr}

\begin{figure}[!htb]
\centering
\includegraphics[width=1\linewidth]{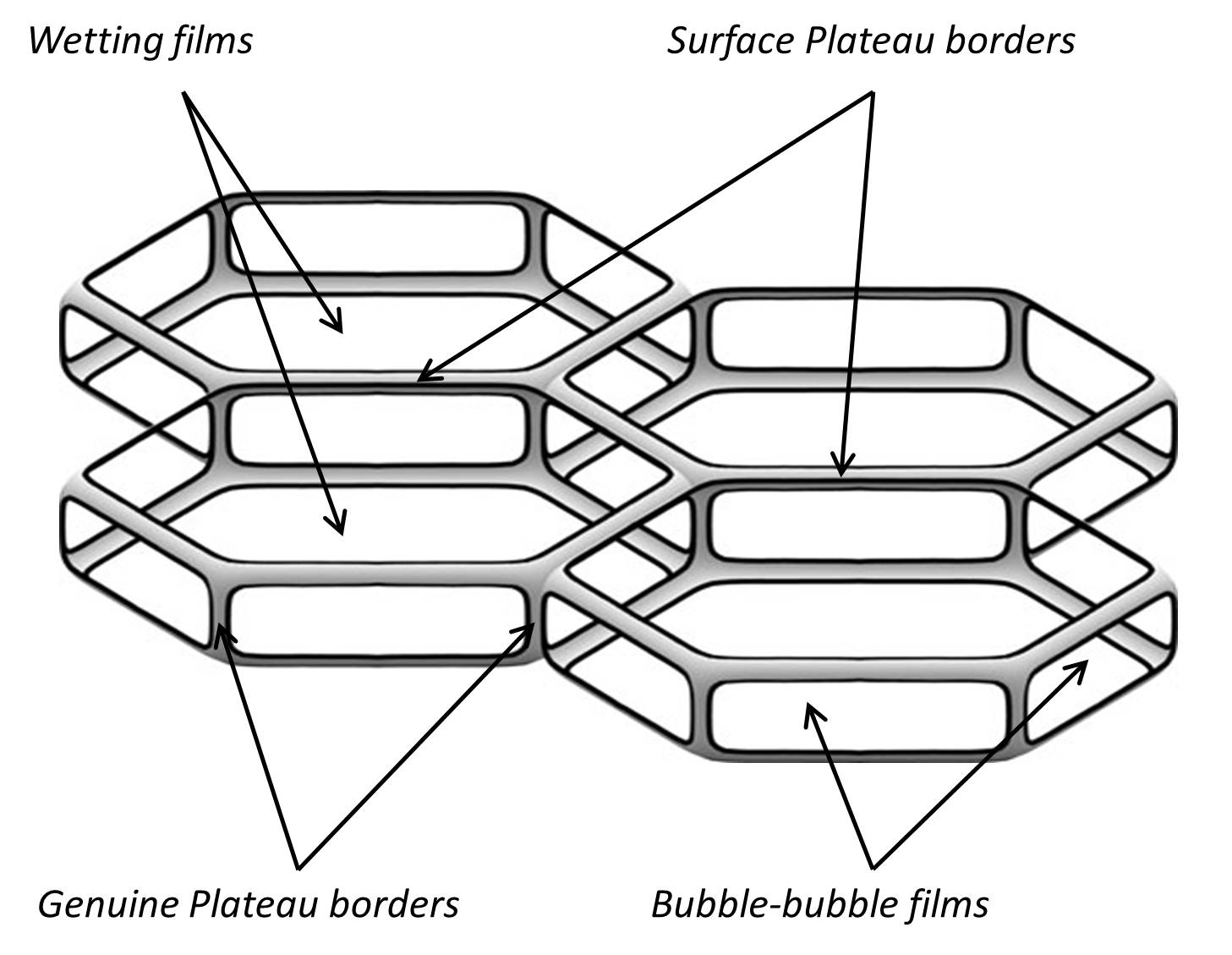}
\caption{A structure of quasi-2D foam squeezed between two glass plates. The picture was obtained from \cite{Gay2011}., with the kind permission of The European Physical Journal (EPJ).}
\label{2Dfoamstructure}
\end{figure}

When a monolayer of bubbles is squeezed between two solid plates in such a manner that each bubble touches both plates a quasi two-dimensional foam is formed (also referred to as a 2D glass-glass (2D GG) foam in a Hele-Shaw cell \cite{Cox2008, Gay2011}, see Figure \ref{2Dfoamstructure}).  Properties of such systems attract significant interest as their local two-dimensional structure can be directly observed which is awkward for classical 3D foams, yet many physical phenomena linked to foam ageing or rheology can be analogously studied. This makes quasi-2D foams an excellent model system.

Quasi-2D foams are  used to study different aspects of  foam rheology \cite{Princen1983, Khan1986, Weaire1989, Cox2004, Katgert2008, Raufaste2009, Costa2013}. Horizontal quasi-2D foams are not significantly affected by drainage as long as the plate spacing is smaller than the capillary length giving the opportunity to investigate coarsening of bubbles \cite{Glazier1987, Stavans1993, Roth2013, Duplat2011}. Coalescence in quasi-2D foams was studied by gently heating them \cite{Burnett1995}, the obtained results were afterwards supported by computer simulations \cite{Hasmy1999}. The dynamics of the topological rearrangements was studied in dry quasi-2D foams \cite{Durand2006}. Instabilities induced by a localized injection of air in quasi-2D foams were studied experimentally and theoretically by Dollet et al \cite{Salem2013, Salem2013b}. Bubble monolayers are also often used to investigate a foam flow in a confined geometry\cite{Jones2011, Raufaste2007}. Another very interesting possible application of quasi-2D foams is  to study more complicated systems such as foamed emulsions (a mixture of bubbles and oil droplets in the water) or foams containing nanoparticles. 

Quasi-2D foams could be even more widely used but often a measure of the liquid content and a proper description of the detailed three-dimensional geometry is required. Despite the ease of observation of the bubble size, a description of the full structure is quite complicated. In addition to the bubble size and liquid content typically used to characterize 3D foams a degree of squeezing (a ratio of thickness of the bubble monolayer to the bubble diameter) plays an important role for quasi-2D foams giving an enormous variety of different bubble shapes. An experimental determination of the exact geometry is hindered: the main source of structural information still remains the computer simulation \cite{Cox2008}.

 Valuable information about the foam structure can be obtained from electrical conductivity measurements. The electrical conductivity is very sensitive to the foam geometry and is now widely used to investigate foams. The measure of electrical conductivity is a powerful tool to determine the foam liquid fraction $\varepsilon$ defined as the ratio between the volume of liquid and the total volume of the foam. For 3D foams the relative conductivity $\sigma$ (being the ratio of the foam $\sigma_{f}$ and the liquid $\sigma_{l}$ conductivities) is found to be primarily a function of the liquid fraction $\varepsilon$ and does not depend on the bubble size \cite{Feitosa2005conductivityfoam}. Theories describing the exact form of this function are well elaborated for three-dimensional foams \cite{Weaire2005} in two limiting cases of "dry" and "wet" foams. The dry foam limit is generally taken as $\varepsilon \lesssim 0.05$, such that the foam can be considered as being composed of polyhedral bubbles whose edges are "decorated" with liquid channels, the so-called Plateau borders \cite{Cantat2013, Weaire2005}. In the wet limit the foam contains enough liquid so that bubbles are nearly spherical. In the dry limit for 3D foams the Plateau borders  may be approximated by straight conductors \cite{Weaire2005}. Taking into account the topology of the foam, considered to be isotropic, Lemlich predicted that \cite{Lemlich1978}

  \begin{equation}
  \sigma=\frac{1}{3}\varepsilon.
  \label{eq 3D Lemlich} 
 \end{equation}
 
 which has been confirmed by numerous experiments \cite{Weaire2005, Cantat2013, Feitosa2005conductivityfoam}. The equation does not take into account the effect of the swollen junctions of Plateau borders. But for dry foams these junctions typically contain negligible amounts of liquid in comparison with the Plateau borders and do not significantly influence the conductivity\cite{Weaire2005}.
In the case of anisotropic dry foams it was shown that Lemlich's limit gives an exact upper bound for the conductivity \cite{Durand2004}.  
 
The well-known Maxwell equation \cite{Maxwell1873} describes conductivity of a media with random spherical insulating inclusions
 
  \begin{equation}
  \sigma=\frac{2\varepsilon}{3-\varepsilon}.
  \label{eq 3D Maxwell} 
  \end{equation}
  This equation is correct in a very-wet limit ($\varepsilon \gtrsim 0.36$) for isolated spherical bubbles in a liquid.
  
   In between the dry and the wet limits different semi-empirical relations have been suggested \cite{Feitosa2005conductivityfoam, Weaire2005, Lemlich1985} in order to smoothly link the limits. 
 
 The above-mentioned limits can be easily rewritten in a two-dimensional space for a hypothetical "true" 2D foam. Such foams do not exist in reality but represent a useful theoretical model. The 2D Lemlich's formula gives \cite{Lemlich1978}
 
 \begin{equation}
\sigma_{2D}=\frac{1}{2} \varepsilon_{2D}.
\label{eq 2D Lemlich} 
\end{equation}

 where $\sigma_{2D}$ and $\varepsilon_{2D}$ are two-dimensional conductivity and liquid fraction, respectively.
 The 2D Maxwell's equation becomes \cite{Maxwell1873}
 
 \begin{equation}
	\sigma_{2D}=\frac{\varepsilon_{2D}}{2-\varepsilon_{2D}}.
	\label{eq 2D Maxwell}
\end{equation}

  However the above-mentioned 2D equations cannot be directly applied to quasi-2D foams. To properly describe the electrical conductivity of quasi-2D foams their real three-dimensional geometry should be taken into account. As far as we know it has never been done before. This is surprising, considering that conductivity measurements may provide an easy solution to the challenge of determining the liquid content of quasi-2D foams which are used by many researchers to access foam properties at the bubble scale.

In the present work we show a model describing the quasi-2D foam geometry and propose geometrical parameters which can be extracted from experimental data simply using photographs. Using the example of ordered, monodisperse foams we discuss how the electrical conductivity can be related to these parameters and how it can help us to investigate the geometry of quasi-2D foams.

\section{Materials and methods}
\label{sec:MaterialsAndMethods}
\begin{figure}[!htbp]
\centering
\begin{minipage}[b]{1\linewidth}
	\includegraphics[width=1\linewidth]{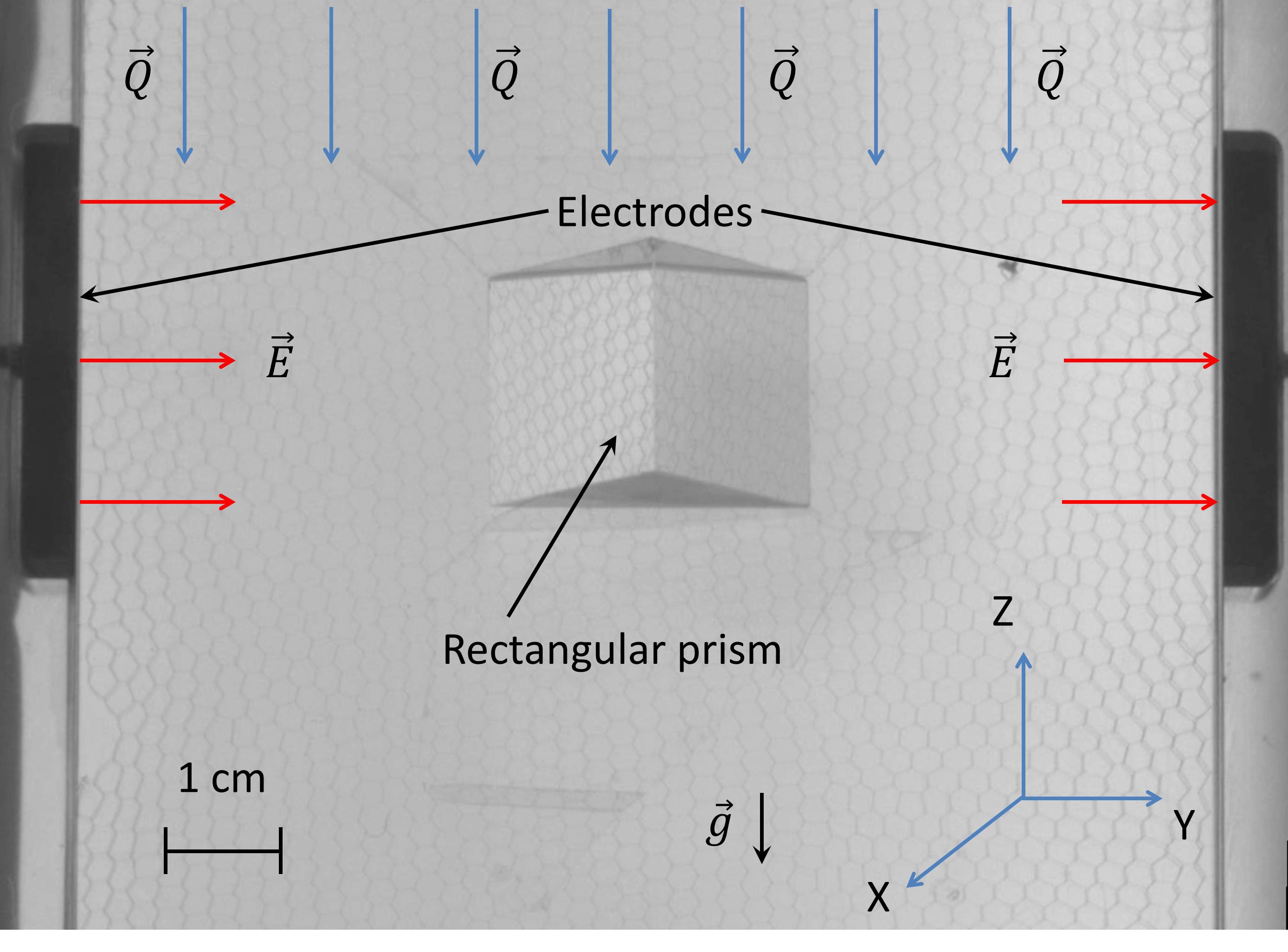}
		\caption{Photograph of the experimental Hele-Shaw cell. A pair of electrodes allows to measure the electrical conductivity. A rectangular prism in the center of the cell is used to take the high-resolution photos of the surface Plateau border network. Foaming solution is injected at constant flow rate $Q$ at the top of the cell to control the liquid fraction of the foam.}
	\label{fig:2D}
	\end{minipage}
	\hfill
	\begin{minipage}[b]{1\linewidth} 
	\includegraphics[width=1\linewidth]{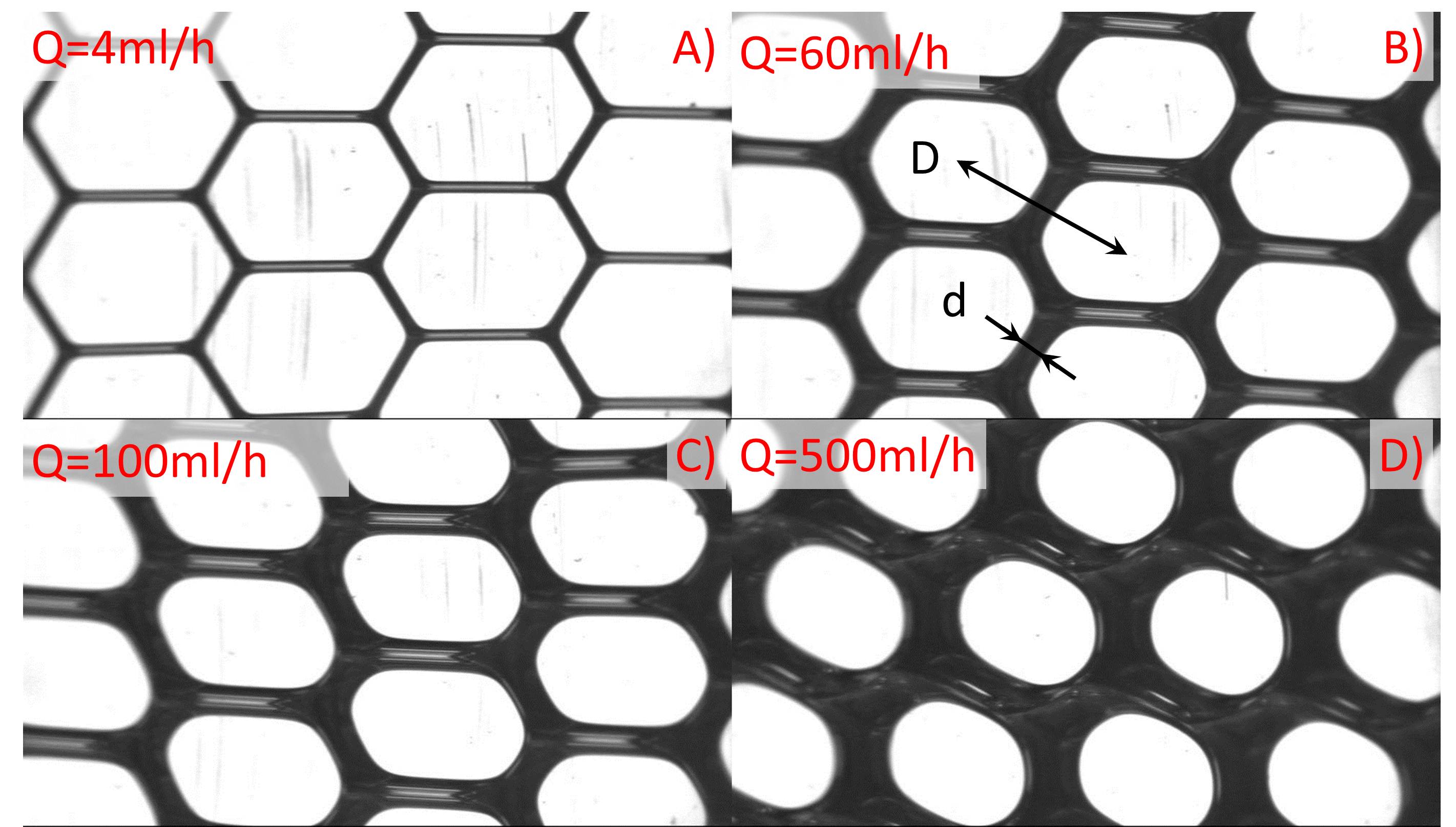}
\caption{High-resolution photographs of vertical quasi-2D foams corresponding to different liquid flow rates: A) $Q=4$ ml/h B) $Q=60$ ml/h C) $Q=100$ ml/h D) $Q=500$ ml/h. The thickness $d$ of the surface Plateau border  and the bubble diameter $D$ are indicated.}
\label{Montagewetanddryfoam}
\end{minipage}

\end{figure}

In our experiments a vertical home-made Hele-Shaw cell consisting of two plexiglas plates with dimensions 10 cm $\times$ 50 cm is used (see Figure \ref{fig:2D}). The distance H between the two plates can be slightly varied but it is typically about 2 mm.  The foam is produced by blowing nitrogen through a single orifice into a solution of Sodium Dodecyl Sulfate (SDS) purchased from Sigma-Aldrich. The surfactant concentration is kept constant at 12 g/L (approximately 5 times the critical micelle concentration) to avoid any surfactant depletion during the generation of the foam. Three pairs of electrodes measure the conductivity at different foam heights. Before each experiment the cell is filled with the foaming solution to have a reference conductivity $\sigma_{l}$. To avoid electrolysis of the foam an alternating current is used with a frequency of 1 kHz and a voltage of 1 V. At the chosen frequency the capacitance of the foam can be neglected and the active resistance can be directly measured \cite{Cantat2013}.

To vary the liquid fraction over a wide range, the experiments are performed in a forced drainage regime: foaming solution is added from the top of the foam at a constant flow rate $Q$. The liquid fraction can be adjusted using different liquid flow rates: a higher flow rate results in a higher liquid fraction \cite{Weaire2005}. Such a regime significantly simplifies our investigation  providing us with a liquid fraction which is not only constant with time but also throughout the entire foam. A steady-state, defined by constant conductivity, is reached before each measurement.

The thickness of the wetting films $h_{wf}$ (see Figure \ref{picture Maxwell}) between the bubbles and the confining plates is calculated from the reflected light spectrum measured by an USB 400 Ocean Optics spectrometer. It was found that in our experiments $h_{wf}$  shows negligible dependence on the flow rate of the liquid and is constant at $3\pm 1$ $\mu m$ within the experimental error. Assuming that the contribution of the wetting film conductivity is directly proportional to this thickness we subtract it systematically from the experimentally measured value  of the relative conductivity $\sigma_{m}$ to have a pure signal from the foam

\begin{equation}
	\sigma=\sigma_{m}-\frac{2h_{wf}}{H}.
\end{equation}

High-resolution photographs of the foam are made with a CCD camera equipped with a telecentric lens through a rectangular prism glued to the outside of the container wall (see  Figure \ref{Montagewetanddryfoam}). This technique was first proposed by Garrett et al. (cited in \cite{Mukherjee1995}). Slight deviations in the path of light reflected by the curved interfaces of the Plateau borders avoid that this light enters the camera. The full surface Plateau border network appears therefore in black. The described optical configuration provides us with reliable information on the foam structure. Each image is treated with ImageJ software to get an average distance $D$ between the centres of the adjacent bubbles and an average surface Plateau border thickness $d$ as shown in Figure \ref{Montagewetanddryfoam}. In the case of dry foams $D$ corresponds to the bubble size. Also a fraction of the surface covered by water $\varepsilon_{surface}$ is calculated for each image.

\section{Geometry of quasi-2D foams}
\label{sec:quasi2DFoamGeometry}

\begin{figure}[!htb]
\centering
\includegraphics[width=1\linewidth]{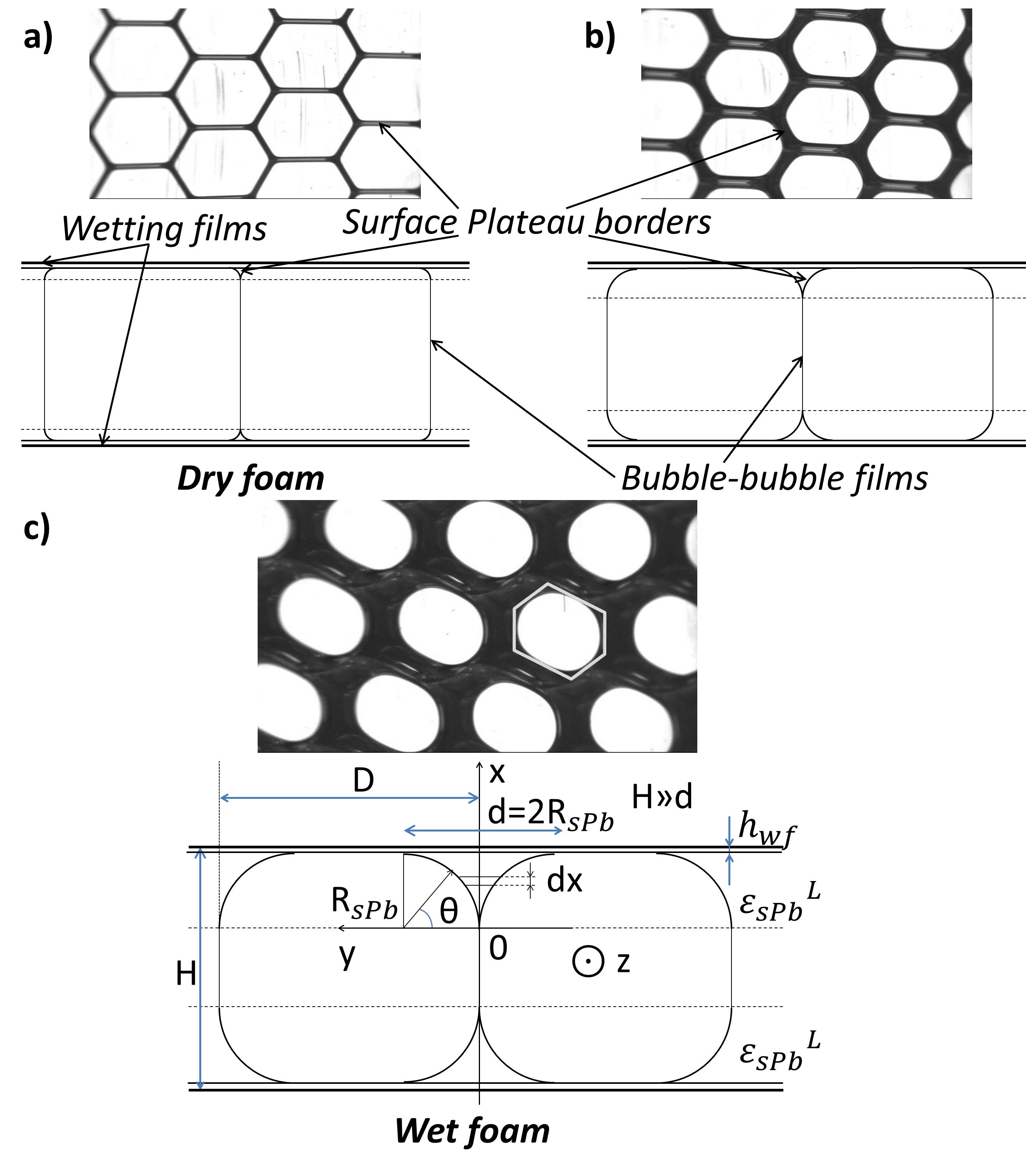}
\caption{A top view and a cross-sectional view perpendicular to the walls for different liquid fractions. A transition from dry to wet limit is shown. The situation depicted here corresponds to the case when the surface Plateau border radius of curvature $R_{sPb}$ is much smaller than the distance between the plates $H$. A hexagon circumscribing a single bubble is shown by a solid grey line.}
\label{picture Maxwell}
\end{figure}

To describe the  properties of quasi-2D foams detailed information about their three dimensional structure is necessary.

Though our experimental foams are never perfectly ordered we will limit our theoretical discussion to the case of ordered monodisperse foams in equilibrium. They are much easier to model and have been found to be very useful for the description of real foams \cite{Drenckhan2010}. It can easily be  shown with Euler's equation that in the case of a monodisperse ordered foam each bubble has exactly six neighbours (as shown in Figure \ref{2Dfoamstructure}). Each bubble is therefore surrounded by 6 Plateau borders (liquid channels that run across the gap between both solid plates at the junction between three bubbles), 12 surface Plateau borders (liquid channels that run along the solid plates at the junction between two bubbles), 6 films separating bubbles (simply referred to as films)  and 2 wetting films covering the surface of the plates. A junction of three surface Plateau borders and one genuine Plateau border is called a node or a vertex.

 In quasi-2D foams all these components make a contribution to the total liquid content but only the surface Plateau borders and wetting films play an important role for the electrical conductivity. This is because the genuine Plateau borders are perpendicular to the electrical flow. An electrical potential is constant along the genuine Plateau borders and their liquid content has negligible influence on the conductivity. The films separating bubbles are always thin in comparison to the surface Plateau borders, that is why the conductivity through them can be neglected. A contribution of the wetting films can be taken into account as explained in Section \ref{sec:MaterialsAndMethods}. We will therefore pay attention in the further discussion of the quasi-2D foam structure only to the surface Plateau borders. 

Viewed from above, only the surface Plateau borders are visible: they form two  identical hexagonal honey-comb networks. We consider here the case of two completely separated surface Plateau border networks characterized by a radius of curvature $R_{sPb}$ much smaller than the gap between the plates, i.e. $R_{sPb}\ll H$. A corresponding cross-sectional view perpendicular to the walls is shown in Figure \ref{picture Maxwell}. In this case the surface networks can be considered as completely independent which significantly simplifies the theoretical description. Also, in the described geometry $R_{sPb}=d/2$ \cite{Gay2011}, the surface Plateau border radius of  curvature can be easily extracted from the photos. So our discussion is limited only to this case and in the experiment the foam is always  maintained in the above described regime.

The liquid content in the surface Plateau border network $\varepsilon_{sPb}$ can be naturally determined as a ratio of the volume of the surface Plateau borders to the total volume of the foam.
However, defined in this manner the surface Plateau border liquid fraction has an important drawback. It depends on the gap between the glass plates and cannot represent the real state of the surface network. We can virtually increase the gap without any change of surface Plateau border structure. So it can not be used as a parameter characterizing the foam geometry in a unique manner. A value free of these disadvantages is a \textit{layer liquid fraction}, which only considers the layer of height $R_{sPb}$ as shown in Figure \ref{picture Maxwell}. It can be expressed as

\begin{equation}
	\varepsilon_{sPb}^{L}=\varepsilon_{sPb}\frac{H}{d}.
	\label{defenition layer liquid fraction}
	\end{equation}
	
Thus determined the layer liquid fraction reflects the real state of the network.
	
The surface Plateau border network  can be modelled as a stack of infinitely thin slices parallel to the wall (see Figure \ref{picture Maxwell}c). This allows the layer liquid fraction to be determined by integration
 
	\begin{equation}
	\varepsilon_{sPb}^{L}=\frac{2}{d}\int_{0}^{d/2}\varepsilon_{2D}(x)dx,
	\label{eq layer liquid fraction integration}
\end{equation}
where $\varepsilon_{2D}(x)$ is a 2D liquid fraction in a slice at a height $x$. To further simplify the calculations an angle $\theta$ is introduced as shown in Figure \ref{picture Maxwell}c. Then the 2D liquid fraction in a given cut can be determined from simple geometrical arguments and written as

\begin{equation}
	\varepsilon_{2D}(\theta)=1-G\left(1-\frac{d}{D}(1-\cos(\theta))\right)^{2},
	\label{eq 2D liquid fraction vs theta}
\end{equation}

where $G$ is a \textit{bubble shape factor} depending on the shape of the bubble cross-section in a given cut. Mathematically it can be defined as the ratio of the bubble area to the area of a circumscribed regular hexagon (see Figure \ref{picture Maxwell}c).  Two important limiting cases for the bubble shape geometry can be distinguished: hexagonal and circular. The first regime is experimentally observed for dry foams (Figure \ref{picture Maxwell}a) while the second one is obtained for wet foams (Figure \ref{picture Maxwell}c). It can be easily shown that for circular bubbles $G=\frac{\pi}{2\sqrt{3}}\approx0.906$, while for hexagonally-shape bubbles $G=1$.

\begin{figure}[!htbp]
\centering

\includegraphics[width=1\linewidth]{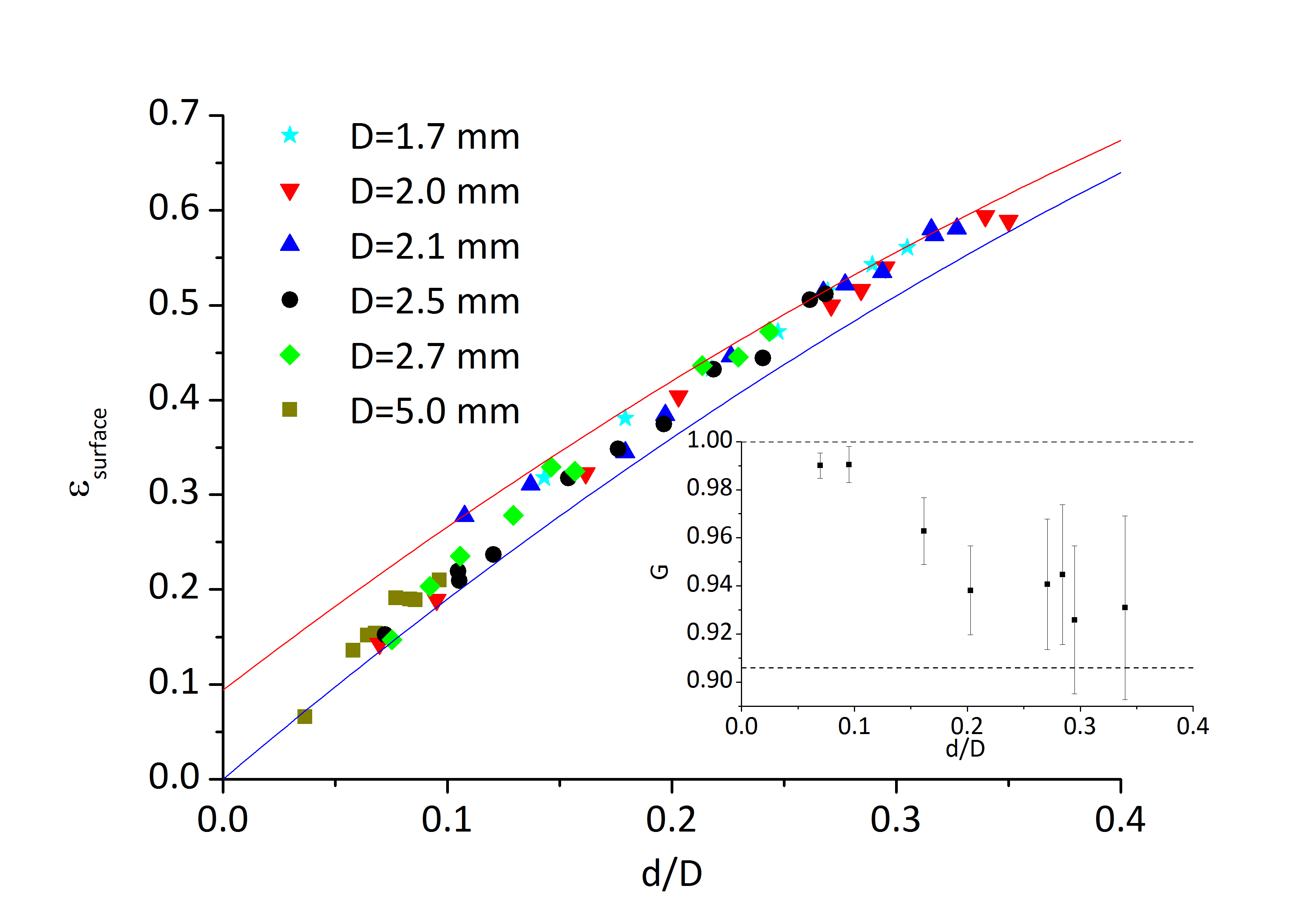}
\caption{Wetted fraction of the wall $\varepsilon_{surface}$ vs $d/D$. Blue and red lines correspond to the prediction of the equation \ref{eq area liquid fraction vs xi} in the dry ($G=1$) and wet ($G=\pi/(2\sqrt{3})$) limits respectively. Inset: $G$ vs $d/D$ dependence (colour online).}
\label{wettedareavsdD}
\end{figure}
	
	\begin{figure}[!htbp]
	\includegraphics[width=1\linewidth]{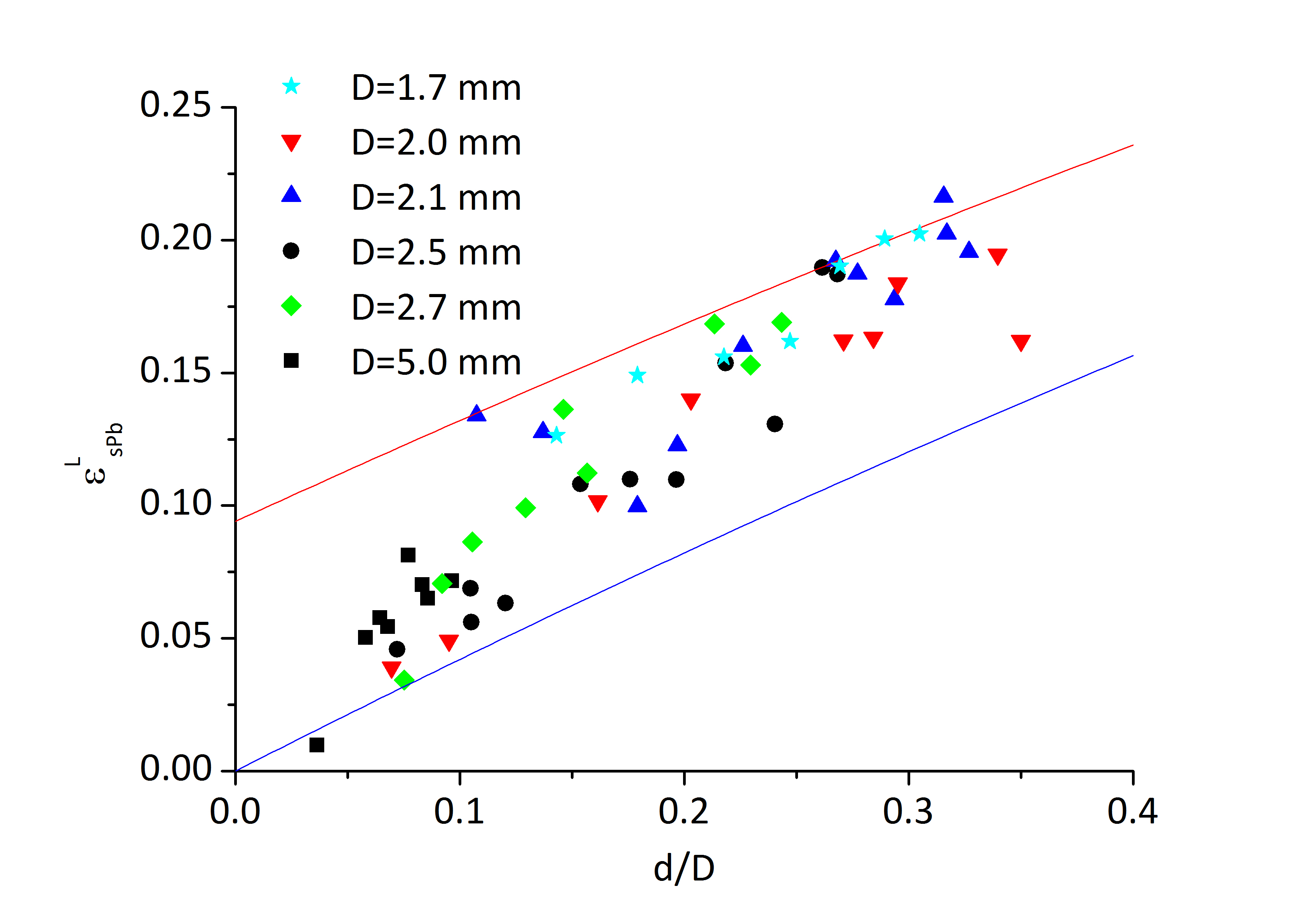}
\caption{The layer liquid fraction $\varepsilon_{sPb}^{L}$ as a function of $d/D$. Blue and red lines correspond to the prediction of the equation (\ref{layer liquid fraction vs xi}) in the dry ($G=1$) and wet ($G=\pi/(2\sqrt{3})$) limits  respectively (colour online).}
\label{layerlfrvsdD}
	
\end{figure}
	
For $\theta=\pi /2$ the 2D liquid fraction corresponds to a wetted fraction at the wall $\varepsilon_{surface}$ which can be extracted from the foam images such as shown in Figure \ref{Montagewetanddryfoam}. Inserting $\theta = \pi /2$ into the Equation (\ref{eq 2D liquid fraction vs theta}) gives

 \begin{equation}
	\varepsilon_{surface}=\varepsilon_{2D}\left(\frac{\pi}{2}\right)=1-G+2G \frac{d}{D}-G\left(\frac{d}{D}\right)^{2}.
	\label{eq area liquid fraction vs xi}
\end{equation}

Equation (\ref{eq area liquid fraction vs xi}) gives us a way to evaluate $G$ directly from the experimental data 

 \begin{equation}
	G=\frac{1-\varepsilon_{surface}}{(1-\frac{d}{D})^{2}}.
	\label{eq G}
\end{equation}

Figure \ref{wettedareavsdD} shows the experimentally measured wetted fraction and $G$ vs $d/D$. One can see that for small $d/D$ foam can be considered as being dry ($G\approx1$), while for $d/D$ higher than 0.1  a transition to the wet limit can be clearly observed ($G\approx0.906$). For small $d/D$ and consequently dry foams $G$ is close to 1 but with an increase of $d/D$ it decreases. As it can be observed from photographs in Figure \ref{Montagewetanddryfoam} bubbles get rounder as $G$ increases.

 The evaluation of the integral in Equation (\ref{eq layer liquid fraction integration}) in combination with Equation (\ref{eq 2D liquid fraction vs theta}) finally gives an expression for the layer liquid fraction
 
\begin{equation}
	\varepsilon_{sPb}^{L}=(1-G)+\left(2-\frac{\pi}{2}\right)G\frac{d}{D} +\left(\frac{\pi}{2}-\frac{5}{3}\right)G\left(\frac{d}{D}\right)^{2}.
	\label{layer liquid fraction vs xi}
\end{equation}

The third term in Equation (\ref{layer liquid fraction vs xi}) is always small and can be neglected. So the layer liquid fraction can be expressed as a linear dependence on $d/D$

\begin{equation}
	\varepsilon_{sPb}^{L}=(1-G)+\left(2-\frac{\pi}{2}\right)G\frac{d}{D}.
	\label{layer liquid fraction vs xi linear}
\end{equation}

 Equations (\ref{eq G}) and (\ref{layer liquid fraction vs xi}) allow us to calculate the layer liquid fraction of the foam from the experimental data. In Figure \ref{layerlfrvsdD} the values calculated from the data are shown as a function of $d/D$. One can see that the layer liquid fraction goes from the dry to the wet limit and reaches relatively high values (more than 20\%).

  \begin{figure}[!htbp]
\centering
	\includegraphics[width=1\linewidth]{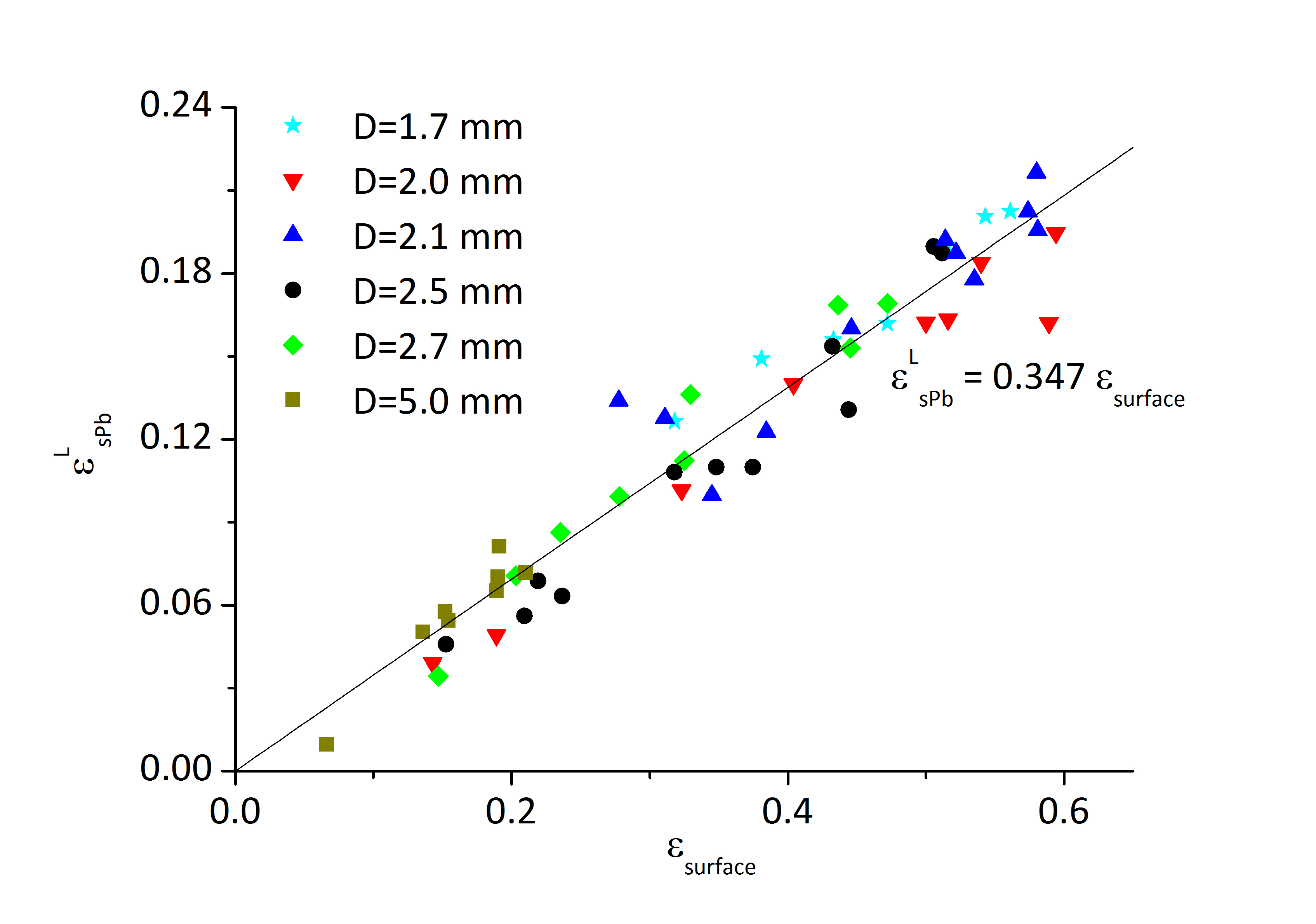}
\caption{The layer liquid fraction $\varepsilon_{sPb}^{L}$ vs the surface liquid fraction $\varepsilon_{surface}$}
\label{lfrvswettedfr}
\end{figure}

  In the presented section three geometrical parameters were introduced to describe the quasi-2D foam geometry: the bubble shape factor $G$, the ratio between the surface Plateau border thickness and the bubble separation $d/D$ and the surface liquid fraction $\varepsilon_{surface}$. Two of them, namely $d/D$ and $\varepsilon_{surface}$, can be directly measured from the photos and the third one ($G$) can be calculated from Equation (\ref{eq G}). But one can ask if $d/D$ and $\varepsilon_{surface}$ can be changed independently. Our experimental results shown in Figure \ref{wettedareavsdD} indicate that in the axes $d/D$ and $\varepsilon_{surface}$ all data lies on one master curve. It implies that there exists a unique dependence of $\varepsilon_{surface}$ on $d/D$.  The exact form of this dependence should be established in the future from more detailed theoretical considerations or computer simulations. However, here we can offer a polynomial fit of the data
  \begin{equation}
	\varepsilon_{surface}\approx2.38\frac{d}{D}-1.89\left(\frac{d}{D}\right)^{2}.
	\label{surfacelfrvsdD} 
	\end{equation}
  The fact that this dependence does exist means that the layer liquid fraction can be expressed as a function of one of the following variables: $G$, $d/D$ or $\varepsilon_{surface}$. For future experimental work with quasi-2D foams a $\varepsilon_{sPb}^{L}$ vs $\varepsilon_{surface}$ dependence is probably the most interesting, since the surface liquid fraction $\varepsilon_{surface}$ is the easiest parameter to estimate from photos. Combining Equations (\ref{eq G}), (\ref{layer liquid fraction vs xi}) and (\ref{surfacelfrvsdD})   we can provide an approximation for $\varepsilon_{sPb}^{L}(\varepsilon_{surface})$ within the framework of our model
   
    \begin{equation}
	\varepsilon_{sPb}^{L}\approx0.347\varepsilon_{surface}.
	\end{equation}
	
	Such a master curve can be very useful for experimentalists as it allows to make a fast and reliable estimation the quantity of liquid in the network of the surface Plateau borders from the photos. 
  
\section{Electrical conductivity}
\label{sec:ElectricalConductivity}

First of all, one should state that in a quasi-2D foam an electrical current can pass only through the network of surface Plateau borders and through the wetting films. Genuine Plateau borders are perpendicular to the electric field and do not contribute to the conductivity. Also in most quasi-2D foams the liquid fraction in the bubble separating films is sufficiently small. It can be estimated that the thin films contain about $10^{-3}$ of the total amount of liquid \cite{Saulnier2015}, so we can neglect the conductivity through them. The addition of the wetting film conductivity can be subtracted as it is explained in Section \ref{sec:MaterialsAndMethods} to have a pure signal from the surface Plateau border networks. We do it systematically in all presented data.

We consider the foam as two independent networks of surface Plateau borders as discussed in Section \ref{sec:quasi2DFoamGeometry}. It is useful to introduce a relative conductivity corresponding to one surface layer in exactly the same manner as we have already introduced the layer liquid fraction (Eqn. (\ref{defenition layer liquid fraction}))

\begin{equation}
\sigma_{sPb}^{L}=\sigma \frac{H}{d}.	
\end{equation}
Determined in the described way \textit{the layer conductivity} and \textit{the layer liquid fraction} do not depend on the plate spacing and represent an actual physical state of the surface networks.

We can apply the same approach used in Section \ref{sec:quasi2DFoamGeometry} to calculate the liquid fraction. A single surface plateau border network can be represented as a parallel connection of infinitely thin conductive slices as shown in Figure \ref{picture Maxwell}. Then the relative conductivity of the whole layer can be calculated by integration

\begin{equation}
	\sigma_{sPb}^{L}=\frac{2}{d}\int_{0}^{d/2}\sigma_{2D}(x)dx.
	\label{eq layer conductivity integration}
\end{equation}
To perform an integration a link between the 2D layer conductivity 	$\sigma_{2D}$ and the 2D liquid fraction $\varepsilon_{2D}$ should be established.

By analogy with a 3D foam, two limiting cases for quasi-2D foams can be considered. In the limit of a dry foam each surface Plateau border can be considered as a thin straight conductor of a constant cross-sectional area and a resistance per unit length. Then the 2D Lemlich's formula (\ref{eq 2D Lemlich}) can be used. Taking into account Equation (\ref{layer liquid fraction vs xi}) the conductivity in the dry limit ($G\rightarrow1, d/D \rightarrow0$) can be written as

\begin{equation}
\sigma_{sPb}^{L}=\left(1-\frac{\pi}{4}\right)\frac{d}{D} .
\label{eq conductivity vs d/D Lemlich}
\end{equation}

An alternative way to calculate the layer conductivity of the surface Plateau border networks is to apply the 2D Maxwell equation (\ref{eq 2D Maxwell}) linking a two-dimensional conductivity with the two-dimensional liquid fraction.

 Using Equation (\ref{eq 2D liquid fraction vs theta}) and  performing a change of variable $x=\frac{d}{2}\sin\theta, \theta\in\left[0,\pi/2\right]$ the layer electrical conductivity can be represented as a function of $d/D$ and $G$
\begin{equation}
	\sigma_{sPb}^{L}\left(\frac{d}{D}, G\right)=-1+2\int_{0}^{\pi/2}\frac{\cos(\theta)}{1+G(1-\frac{d}{D}(1-\cos(\theta)))^{2}}d\theta.
	\label{eq conductivity vs d/D}
\end{equation}

\begin{figure}[h]
\centering

\includegraphics[width=1\linewidth]{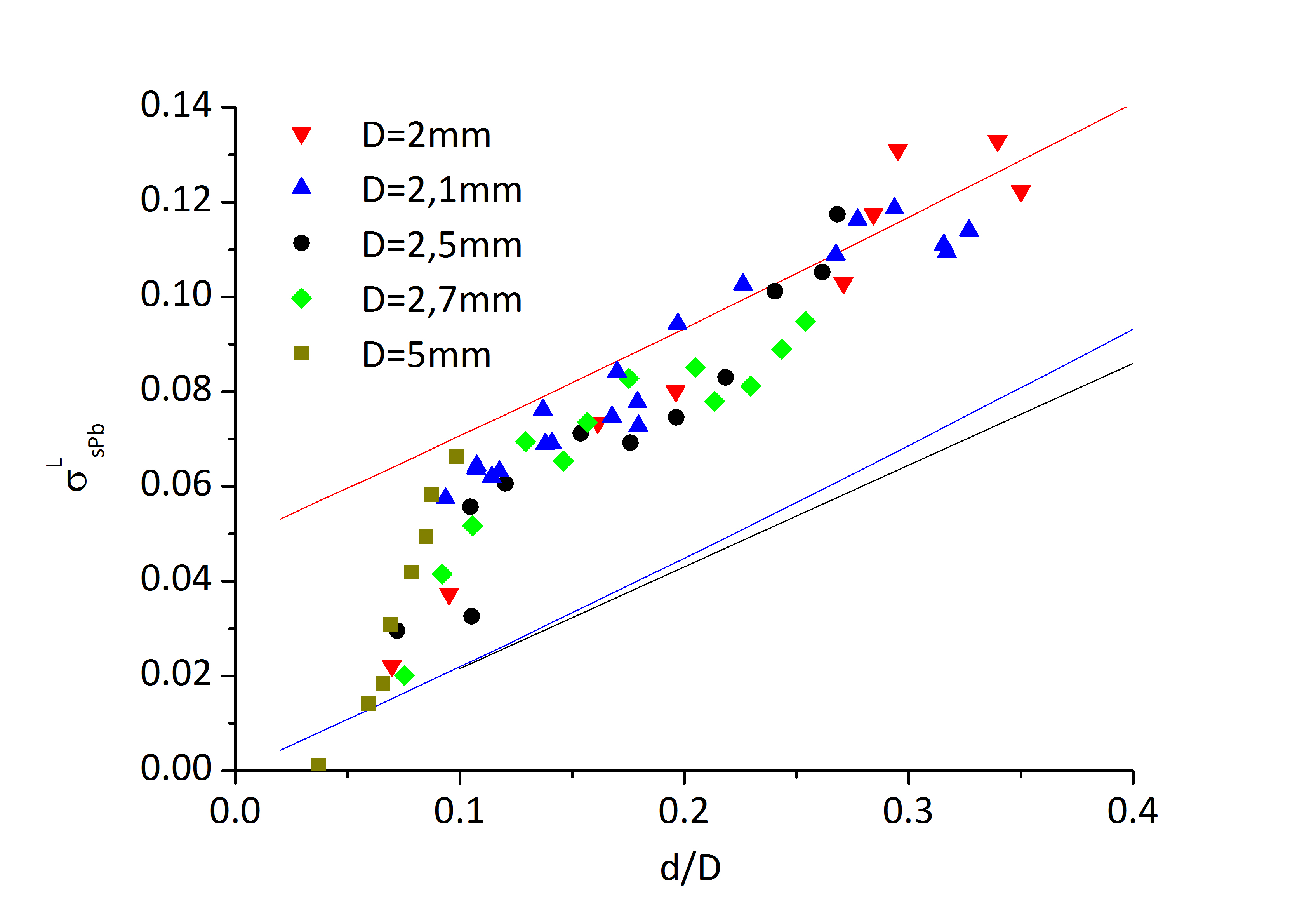}
\caption{The layer electrical conductivity $\sigma_{sPb}^{L}$ vs $d/D$ measured for different bubble sizes and liquid flow rates. Blue and red lines correspond to the prediction of the Maxwell model (Eqn. \ref{eq conductivity vs d/D}) in the dry ($G=1$) and wet ($G=\pi/(2\sqrt{3})$) limits respectively. The black line corresponds to Lemlich's model (Eqn. \ref{eq conductivity vs d/D Lemlich}) (colour online).}
\label{FigureMaxwell-sim}
\end{figure}

The integral of the Equation (\ref{eq conductivity vs d/D}) was numerically evaluated for different values of $d/D$. The results are plotted in Figure (\ref{FigureMaxwell-sim}) for two limiting values of $G$. One can see that for the hexagonal bubbles Maxwell's approach gives a result very close to Lemlich's one for sufficiently dry quasi-2D foams.

 Also an enormous difference between the dry and the wet limits can be observed allowing us to anticipate that electrical conductivity can be used as a sensitive instrument to explore the three-dimensional structure of quasi-2D foams.

\begin{figure}[h]
	
	\includegraphics[width=1\linewidth]{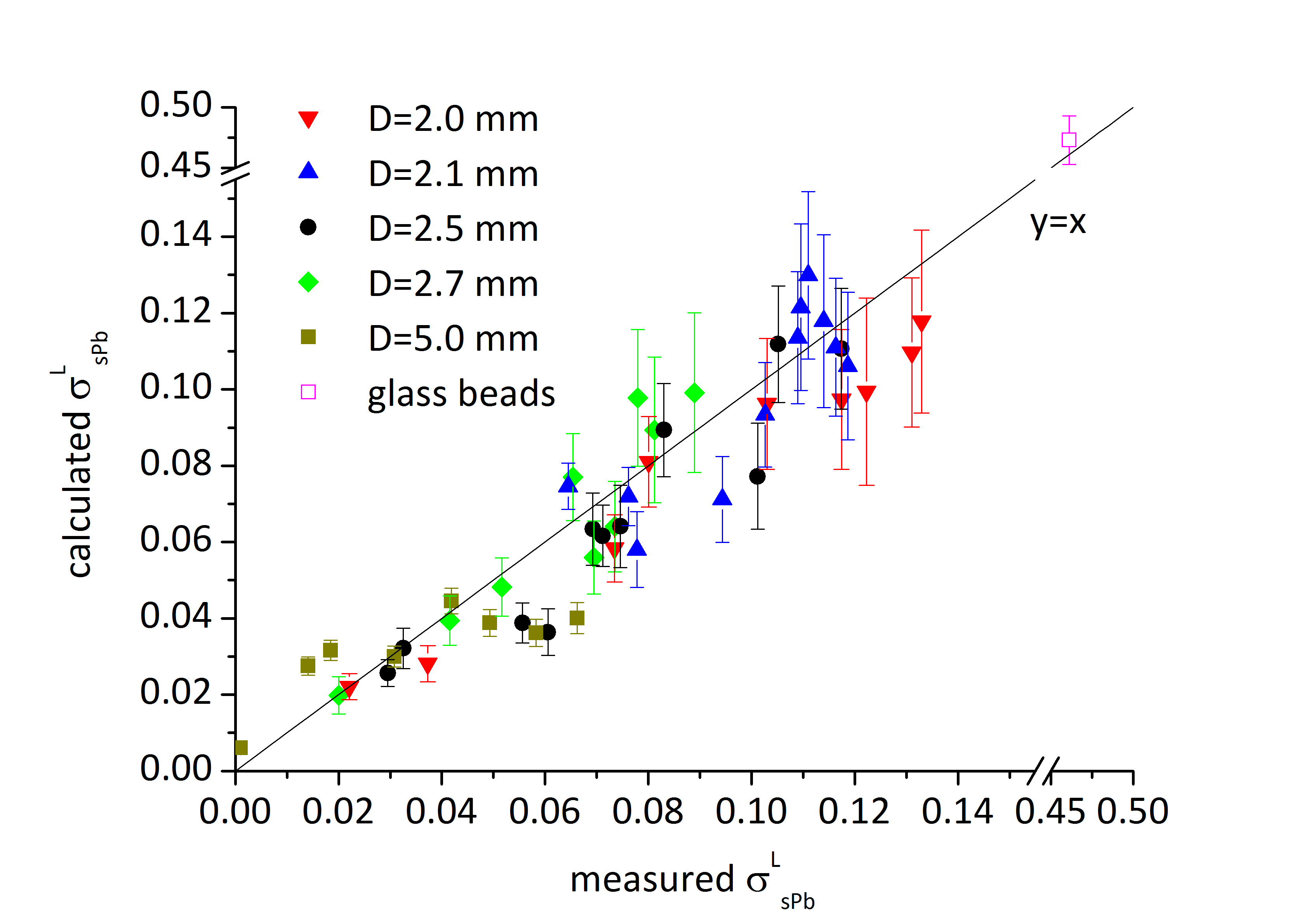}
	\caption{The  layer electrical conductivity  $\sigma_{sPb}^{L}$ calculated by Equation (\ref{eq conductivity vs d/D approx}) using the geometrical parameters extracted from photos vs the experimentally measured values (colour online).}
\label{calculatedelcondvsmeasured}
	\end{figure}

Experimental data for the layer electrical conductivity are represented in Figure \ref{FigureMaxwell-sim} for different liquid fractions and bubble sizes. Except for very dry foams the error bars normally do not exceed a few percent and are comparable with the symbol sizes.  For dry foams ($d/D<0.07$) a certain uncertainty in a wetting film thickness estimation can significantly influence the accuracy of electrical conductivity measurements.

A clear transition between the dry and the wet limit is visible at $d/D\approx0.1$. Such behaviour can be attributed to the transition of the foam structure from the dry to the wet limit. This can be explained by calculating the minimal surface area. For dry foam only a hexagonal structure is possible: to curve bubbles and make them circular a certain amount of liquid is necessary. But  as long as it is possible to make circular bubbles they will always have a smaller area and consequently a lower energy than the hexagonal ones. A geometrical calculation shows that the minimum liquid fraction required to make circular bubbles is about  0.094. This value corresponds to $d/D$ about 0.11 in the hexagonal model. It means that above this value hexagonal bubbles should not exist. In practice it means that after this limit the circularity of bubbles significantly increases and the geometry dramatically changes. This corresponds extremely well with the experimentally observed transition in Figure \ref{FigureMaxwell-sim}.

 The complicated geometry of quasi-2D foams can be taken into account by using the experimentally measured values of $G$. To simplify our calculations a linear approximation of the expression (\ref{eq conductivity vs d/D}) can be used to predict $\sigma_{sPb}^{L}$
 
 \begin{strip}
 \begin{equation}
	\sigma_{sPb}^{L}\left(\frac{d}{D}, G \right)\approx\frac{2}{\sqrt{G(1+G)}}\left( \arctanh \left(\sqrt{\frac{G}{1+G}} \right)-\sqrt{\frac{G}{1+G}} \right)\frac{d}{D}+\frac{1-G}{1+G}.
	\label{eq conductivity vs d/D approx}
\end{equation} 
\end{strip}

For the dry limit this approximation gives
\begin{equation}
	\sigma_{sPb}^{L}\left(\frac{d}{D}\right)\approx 0.246\frac{d}{D} \ \text{(dry limit)}
	\label{eq conductivity vs d/D approx dry limit}
\end{equation}

which agrees well with Lemlich's limit

\begin{equation}
	\sigma_{sPb}^{L}\left(\frac{d}{D}\right)\approx 0.215\frac{d}{D}.
	\label{eq conductivity vs d/D approx dry limit Lemlich}
\end{equation}

In the wet limit we get
\begin{equation}
	\sigma_{sPb}^{L}\left(\frac{d}{D}\right)\approx 0.240\frac{d}{D}+0.049 \ \text{(wet limit)}.
	\label{eq conductivity vs d/D approx wet limit}
\end{equation}

 The values of $\sigma_{sPb}^{L}$ calculated by Equation (\ref{eq conductivity vs d/D approx}) vs the experimentally measured ones are presented in Figure \ref{calculatedelcondvsmeasured}. The values of G are calculated from photos using Equation (\ref{eq G}). The results are in a good agreement with the experimental data and confirm our theoretical assumptions. The obtained approximations (\ref{eq conductivity vs d/D approx dry limit}) and (\ref{eq conductivity vs d/D approx wet limit}) can be very useful for future experimental work since they allow to estimate the electrical conductivity  from photo treatment or to calculate the surface Plateau border thickness from the known conductivity data in two limiting cases of very dry and very wet foams. Along with the equation (\ref{layer liquid fraction vs xi}) it gives a straightforward way to evaluate the surface Plateau border liquid fraction from conductivity measurements in the above-mentioned limits.
 
  \begin{figure}[!htbp]
\centering
	\includegraphics[width=1\linewidth]{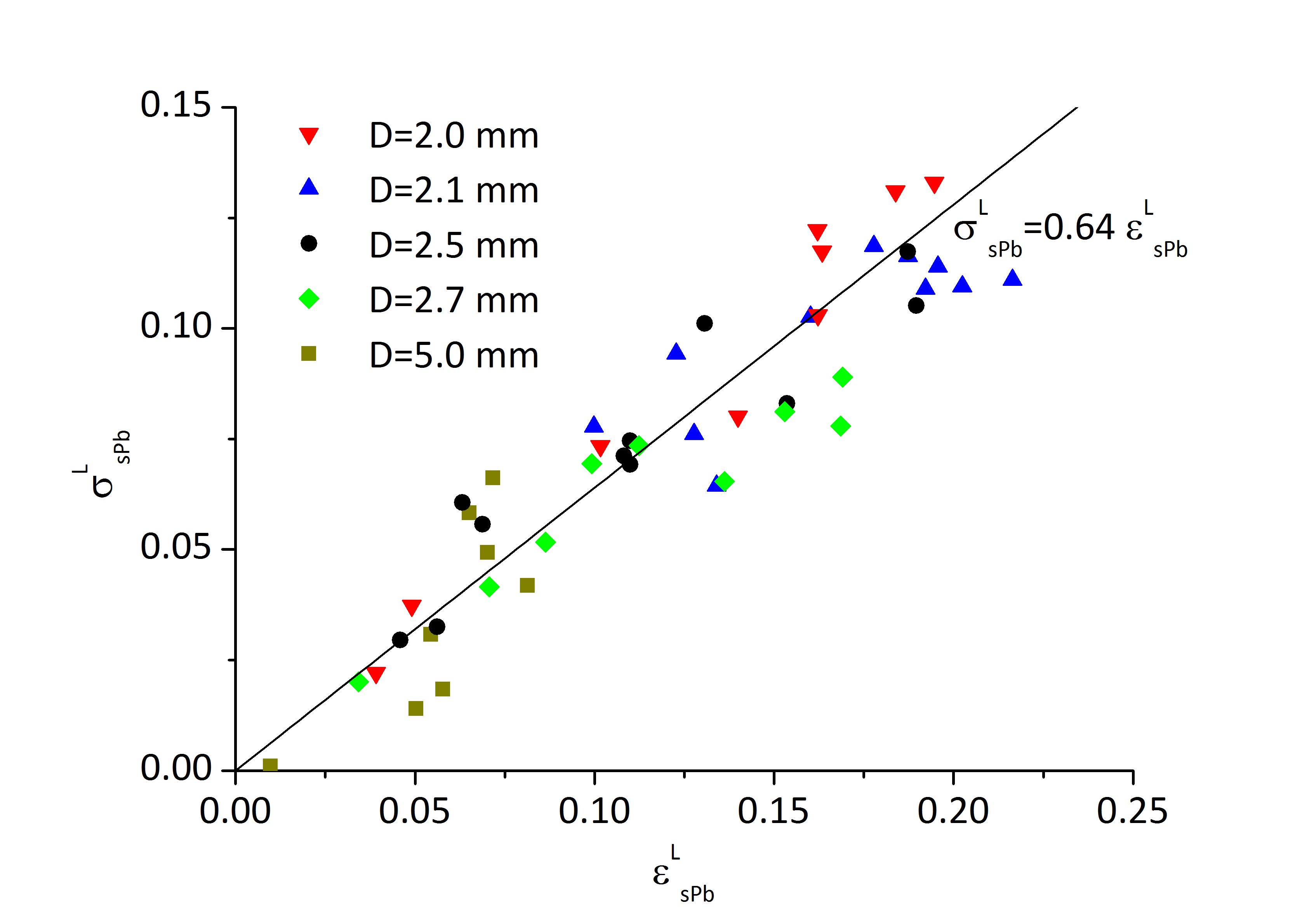}
\caption{The layer electrical conductivity $\sigma_{sPb}^{L}$ vs the layer liquid fraction $\varepsilon_{sPb}^{L}$ for different bubble sizes}
\label{elcondvslfr}
\end{figure}

The dependence of the layer conductivity on the layer liquid fraction is shown in Figure \ref{elcondvslfr}. The data can be approximated by a linear relationship 
\begin{equation}
	\sigma_{sPb}^{L}\approx 0.64\varepsilon_{sPb}^{L}.
\end{equation}
This relationship can be used to rapidly estimate the liquid fraction from the electrical conductivity data.

The low error of layer electrical conductivity measurements and the high sensitivity to the change of the foam structure allows us to determine the geometry of the quasi-2D foam.

To check the applicability of the developed approach for very high liquid fractions we also measured the conductivity of a glass bead monolayer surrounded by the same foaming solution. Such system corresponds to the case $d/D=1$. The measured value of conductivity is in a full agreement with the prediction of our theory as shown in Figure \ref{calculatedelcondvsmeasured}.

\section{Conclusion}
\label{sec:Conclusion}

 In this article we introduced a simple model to describe the geometry of a quasi-2D foam. We used this description to model the electrical conductivity of quasi-2D foams. This model describes well our accompanying experiments over a wide range of liquid fractions. Our experiments show that even foams at low liquid fraction have to be considered as "wet". 

We hope that this work can help in suggesting new approaches for the characterisation of foam properties. In particular, it should prove useful in the in-situ characterisation of foam flow in the presence of walls and in confining geometries, such as microfluidic applications \cite{Marmottant2009, Huerre2014}. 

The reader should also keep in mind that our models are equally valid for liquid/liquid foams, i.e. emulsions, if a non-conduction dispersed phase is used.

\section{Acknowledgments}
\label{sec:Acknowledgments}

 We would like to thank Teclis for help in design and construction of the experimental cell. We acknowledge a financial support from the French space agency CNES. We acknowledge funding from the European Research Council via an ERC grant agreement 307280-POMCAPS.

\newpage
\bibliography{mybiblio}
\end{document}